\documentstyle[epsfig,subfig,amsmath]{elsart}
\newcommand{\newc}{\newcommand}
\newc{\lra}{\leftrightarrow}
\newc{\beq}{\begin{equation}}
\newc{\eeq}{\end{equation}}
\newc{\barr}{\begin{eqnarray}}
\newc{\earr}{\end{eqnarray}}

\newcommand{\beqa}{\begin{eqnarray}}
\newcommand{\eeqa}{\end{eqnarray}}
\newcommand{\bdm}{\begin{displaymath}}
\newcommand{\edm}{\end{displaymath}}

\renewcommand{\thefootnote}{\#\arabic{footnote}}
\renewcommand{\theequation}{\thesection.\arabic{equation}}
\renewcommand{\thefigure}{\thesection.\arabic{figure}}
\renewcommand{\thetable}{\thesection.\arabic{table}}

\begin{document}
\date{\today}
\title {Solar neutrinos as background in dark matter searches involving electron detection}
%
%
%
%
%
\author{A. Thomas$^{1}$ and J.D. Vergados$^{1,2}$}
%
\address{$^1$ CoEPP and Centre for the Subatomic Structure of Matter (CSSM), University of Adelaide, Adelaide SA 5005, Australia}
\address{$^2$Theoretical Physics,University of Ioannina, Ioannina, Gr 451 10, Greece.
\\E-mail:Vergados@uoi.gr}
\begin{frontmatter}
\begin{abstract}
In the present work we estimate  the potential background of solar neutrinos on electron detectors. These detectors are considered relevant for detecting light dark matter particles in the MeV region, currently sought by experiments. We find that the copious low energy pp neutrinos are a dangerous background at the energies involved in these experiments, in fact close to the anticipated event rate, while  the more energetic Boron neutrinos are harmless.
\end{abstract}

\end{frontmatter}
\section{Introduction}

The combined MAXIMA-1 \cite{MAXIMA1},\cite{MAXIMA2},\cite{MAXIMA3}, BOOMERANG \cite{BOOMERANG1},\cite{BOOMERANG2}
DASI \cite{DASI02} and COBE/DMR Cosmic Microwave Background (CMB)
observations \cite{COBE} imply that the Universe is flat
\cite{flat01}
and that most of the matter in
the Universe is Dark \cite{SPERGEL},  i.e. exotic. These results have been confirmed and improved
by the recent WMAP  \cite{WMAP06} and Planck \cite{PlanckCP13} data. Combining 
the data of these quite precise measurements one finds:
$$\Omega_b=0.0456 \pm 0.0015, \quad \Omega _{\mbox{{\tiny CDM}}}=0.228 \pm 0.013 , \quad \Omega_{\Lambda}= 0.726 \pm 0.015~$$
(the more  recent Planck data yield a slightly different combination $ \Omega _{\mbox{{\tiny CDM}}}=0.274 \pm 0.020 , \quad \Omega_{\Lambda}= 0.686 \pm 0.020)$. It is worth mentioning that both the WMAP and the Plank observations yield essentially the same value of $\Omega_m h^2$,
  but they differ in the value of $h$, namely $h=0.704\pm0.013$ (WMAP) and $h=0.673\pm0.012$ (Planck).
Since any ``invisible", non-exotic component cannot possibly exceed $40\%$ of the above $ \Omega _{\mbox{{\tiny CDM}}}$
~\cite {Benne}, exotic (non baryonic) matter is required and there is room for cold dark matter candidates or WIMPs (Weakly Interacting Massive Particles).\\
Even though there exists firm indirect evidence for a halo of dark matter
in galaxies from the
observed rotational curves, see e.g. the review \cite{UK01}, it is essential to directly
detect such matter in order to 
unravel the nature of the constituents of dark matter. 
The possibility of direct detection, however, depends on the nature of the dark matter constituents and their interactions.

The WIMP's are  expected  have a velocity distribution with an average velocity which is close to the rotational velocity $\upsilon_0$ of the sun around the galaxy, i.e.  they are completely non relativistic. In fact a Maxwell-Boltzmann distribution leads to an average energy  $\prec T\succ=0.4\times 10^{-6}m c^2$. Thus for GeV WIMPS this average is in the keV regime, not high enough to excite the nucleus, while for WIMPs with the mass of the order of the electron mass the average energy is in the eV region. So WIMPs this light can be detected only by atomic excitations, following the electron recoil. In the present work we will focus on light WIMPs in the MeV region, most likely to be detected in electron recoil experiments.\\ 
In such experiments, like the nuclear measurements first proposed more than 30 years ago \cite{GOODWIT}, one has to face the problem that the process of interest does not have a characteristic feature to distinguish it
from the background. 

There has  been an interest in light WIMPs recently. For example we mention the work of  Ref. \cite{EMMPV12}, where the first direct detection limits on sub-GeV dark matter from XENON10 have recently been obtained. Furthermore detection of electron recoils has also  been considered recently, see , e.g.,  for atomic electrons \cite{EMV12} as well as superconductors \cite{HPZ15} and references there in.
This is encouraging, but based on our experience with standard nuclear recoil experiments to excited states \cite{VerEjSav13}, one has to make sure that the proper kinematics  is used in dealing with bound electrons. Clearly  the binding  energy of the electron plays  the same role as the excitation energy of the nucleus, in determining the small fraction of the WIMP's energy to be transferred to the recoiling system. In fact it is found that the expected event rates are very much suppressed, if one takes into account the binding energy of the electrons \cite{VMEK16}, at least  for atomic electrons, in which case  binding energies less than 1 eV are rare. It is therefore clear that 
light WIMPs are quite different objects in  energy,  mass, flux and detection. Accordingly 
one needs detectors capable of  detecting low energy, light WIMPs in the midst of formidable backgrounds, i.e. detectors  which are completely different from  current WIMP detectors employed for heavy WIMP searches.  In view of the above suppression one has to consider background effects due to the omnipresent and unavoidable solar neutrinos. This issue will be addressed in the present work.

\section{ Elastic Neutrino Electron Scattering}
 For low energy neutrinos the historic process neutrino-electron scattering \cite{HOOFT} \cite{REINES}
 is very useful.
The differential cross section \cite{VogEng} takes the form
\begin{equation}
\frac{d\sigma}{dT}=\left(\frac{d\sigma}{dT}\right)_{weak}+
\left(\frac{d\sigma}{dT}\right)_{EM} \label{elas1a}
\end{equation}
The second term, due to the neutrino magnetic moment, is inversely proportional to the electron
energy and, at the low electron energies of the present set up, may be used in improving
 the current limit of the neutrino magnetic moment by two orders of magnitude. It is not, however,
 the subject of the present study, which is concerned with the investigation of solar neutrinos as a source of background WIMP-electron scattering. The cross section in the rest frame of the initial electron due to the weak interaction alone becomes:
 \begin{eqnarray}
 \left(\frac{d\sigma}{dT}\right)_{weak}&=&\frac{G^2_F m_e}{2 \pi}
 \left [(g_V+g_A)^2 \right .\nonumber \\
&&\left . + (g_V-g_A)^2\left  (1-\frac{T}{E_{\nu}}\right )^2
+ (g_A^2-g_V^2)\frac{m_eT}{E^2_{\nu}}\right ]
 \label{elasw1}
  \end{eqnarray}
 $$g_V=2\sin^2\theta_W+1/2~~ (\nu_e)~,~g_V=2\sin^2\theta_W-1/2~~ (\nu_{\mu},\nu_{\tau})$$
 $$g_A=1/2~~~~ (\nu_e)~~~~,~~~~g_A=-1/2~~~~ (\nu_{\mu},\nu_{\tau})$$
 with $\theta_W$ the Weinberg angle. For antineutrinos $g_A\rightarrow-g_A$. Here T is the electron energy and $E_{\nu}$ is the neutrino energy.

 In the present study, however, we will find it more convenient to express the cross section in dimensionless units, by expressing  both the electron and the neutrino energy in units of the electron mass, namely $T=y m_e$ and $E_{\nu}=x m_e$. Thus
\beq
 \left(\frac{d\sigma}{dy}\right)_{weak}=\frac{G^2_F m^2_e}{2 \pi}\frac{d\tilde{\sigma}(x,y)}{dy}
\eeq
with
\beq
\frac{d\tilde{\sigma}}{dy}(x,y)=\left [\left (g_V+g_A\right )^2 +  \left (g_V-g_A\right )^2 \left(1-\frac{y}{x} \right)^2
+ \left (g_A^2-g_V^2\right )\frac{y}{x^2}\right ]
 \label{elasw}
  \eeq
  The scale is set by the weak interaction:
\beq
 ws=\frac{G^2_F m^2_e}{2 \pi}=2.27\times 10^{-45}~\mbox{cm}^2
\label{Eq:weakscale}
\eeq

The electron energy depends on the  neutrino energy and the
scattering angle and is given by\cite{GiomVer03}:
\beq
y(x)= \frac{2~ (x\cos{\theta})^2}{(1+x)^2-(x
\cos{\theta})^2}
\label{Te}
\eeq
The maximum electron energy is attained when $\xi=\cos{\theta}=1$ and is given as a function of the neutrino energy by
\beq
y_{max}(x)=\frac{2 x^2}{1+2x}
\label{Eq:ymax}
\eeq
On the other hand to produce an electron with energy $y$ the neutrino energy must be at least \cite{GiomVer05}:
\beq
x_{min}(y)=\left (-1+\sqrt{1+\frac{2}{y}}\right )^{-1}=\frac{1}{2}\left (y+\sqrt{y^2+2 y}\right )
\label{Eq:xmin}
\eeq
The last equations ignore the binding energy of the electrons.

If the neutrino emitter is characterized by  a normalized  energy distribution $f_{dis}(x)$ the folded cross section becomes:
 \beq
\left(\frac{d\sigma}{dy}\right)_{av}=\frac{G^2_F m^2_e}{2 \pi}\prec\left(\frac{d \tilde{\sigma}}{dy}\right)\succ
\label{Eq:avedsdy1}
\eeq
with
\beq
\left(\frac{d \tilde{\sigma}}{dy}\right)\succ=\int_{x_{min}(y)}^{xmax}f_{dis}(x)\frac{d \tilde{\sigma}(x,y)}{dy}dx.
\label{Eq:avedsdy2}
\eeq
The total cross section (per electron) is given by
\beq
\sigma_{tot}=\frac{G^2_F m^2_e}{2 \pi}\int_{0}^{y_{max}(x_{max})}dy\int_{x_{min}(y)}^{x_{max}} f_{dis}(x)\frac{d\tilde{\sigma}(x,y)}{dy}dx
\label{Eq:sigmatot}
\eeq
where $y_{max}(x_{max})$ is the maximum possible electron energy (see Eq. \ref{Eq:ymax}).\\
 The above equation must be modified if one wishes to include the fact that the initial electron is bound. For neutrino energies as those of the boron solar neutrinos it has been shown by Gounaris, Paschos and Porfyriadis \cite{GPP04} that  such effects are negligible. In the case of pp neutrinos the effect of the binding will be examined below.
\section{The solar neutrino spectra}
\begin{figure}
\begin{center}
\includegraphics[width=0.8\textwidth]{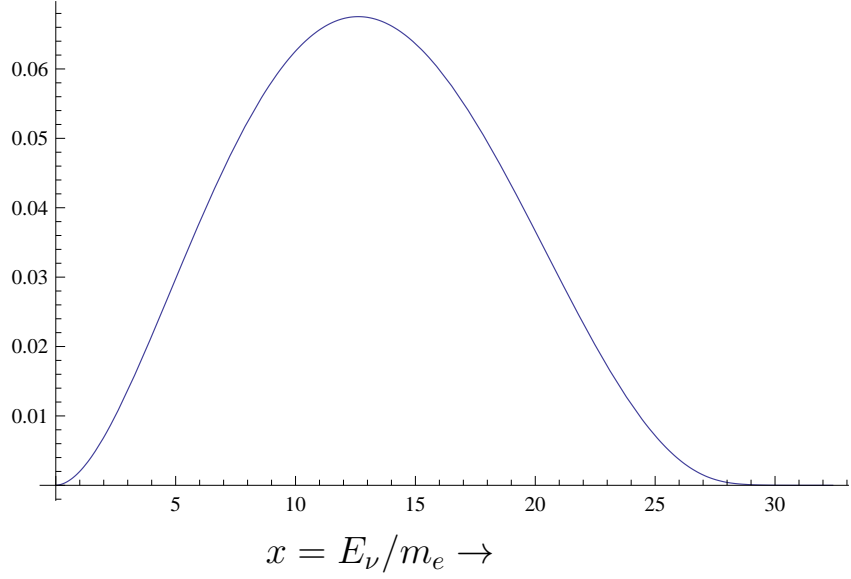}\\
{\hspace{-1.5cm} $x=E_{\nu}/m_e\rightarrow$}
\caption{The normalized boron solar neutrino spectrum given as a function of $x=E_{\nu}/m_e$
 \label{fig:BoronNuSpec}}
\end{center}
\end{figure}

\begin{figure}
\begin{center}
\includegraphics[width=0.8\textwidth]{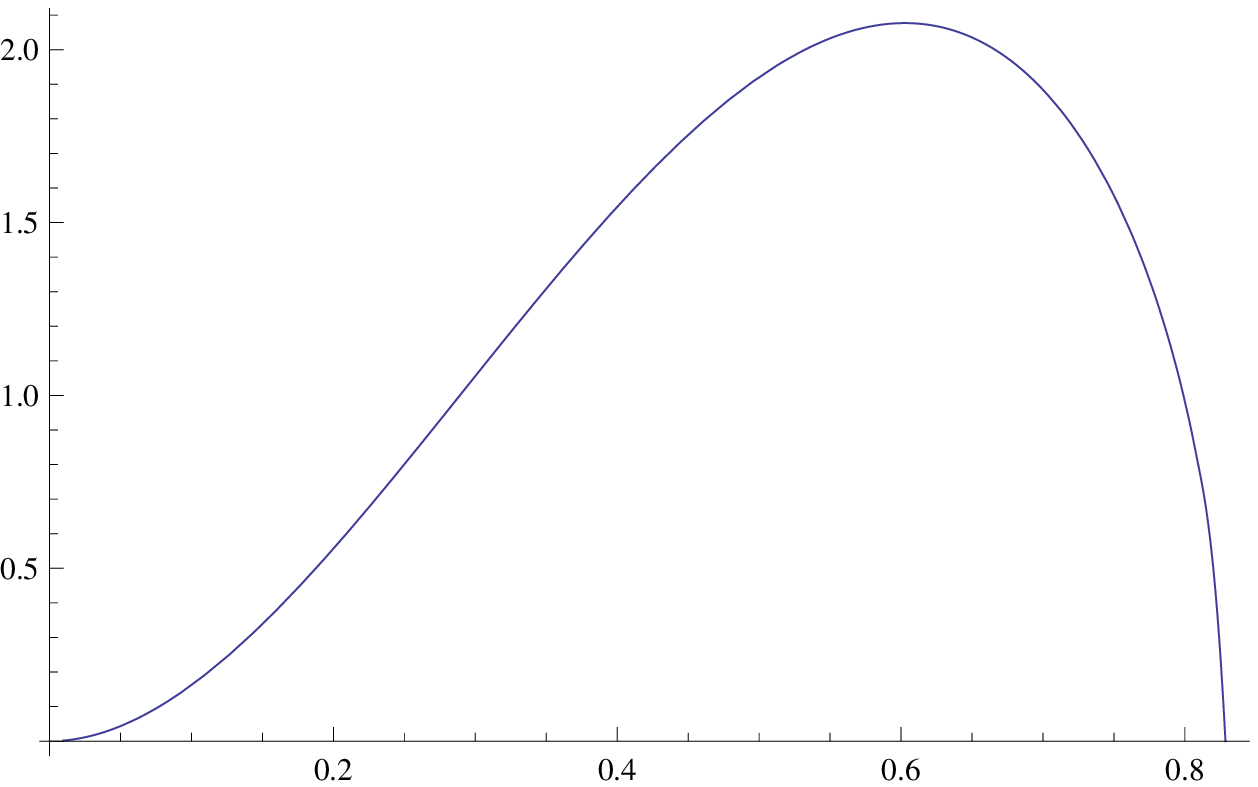}\\
{\hspace{-1.5cm} $x=E_{\nu}/m_e\rightarrow$}
\caption{The normalized boron solar neutrino spectrum given as a function of $x=E_{\nu}/m_e$
 \label{fig:ppNuSpec}}
\end{center}
\end{figure}

The boron solar neutrino spectrum extends from zero to a maximum of 15.5 MeV  \cite{BahSenBas05}and is shown in Fig. \ref{fig:BoronNuSpec}.
The corresponding solar neutrino Flux is
$$\Phi=5.0 \times10^{5}\mbox{cm}^{-2}\mbox{s}^{-1}=1.6\times 10^{13}\mbox{cm}^{-2}\mbox{y}^{-1}$$
For the pp neutrinos the spectrum is exhibited in Fig. \ref{fig:ppNuSpec}.
The average energy is lower, implying a smaller cross section, but the neutrino flux is much larger
$$\Phi=6.2 \times10^{10}\mbox{cm}^{-2}\mbox{s}^{-1}=1.9\times 10^{18}\mbox{cm}^{-2}\mbox{y}^{-1}$$
\section{Differential event rates}
\begin{figure}
\begin{center}
\includegraphics[width=0.8\textwidth]{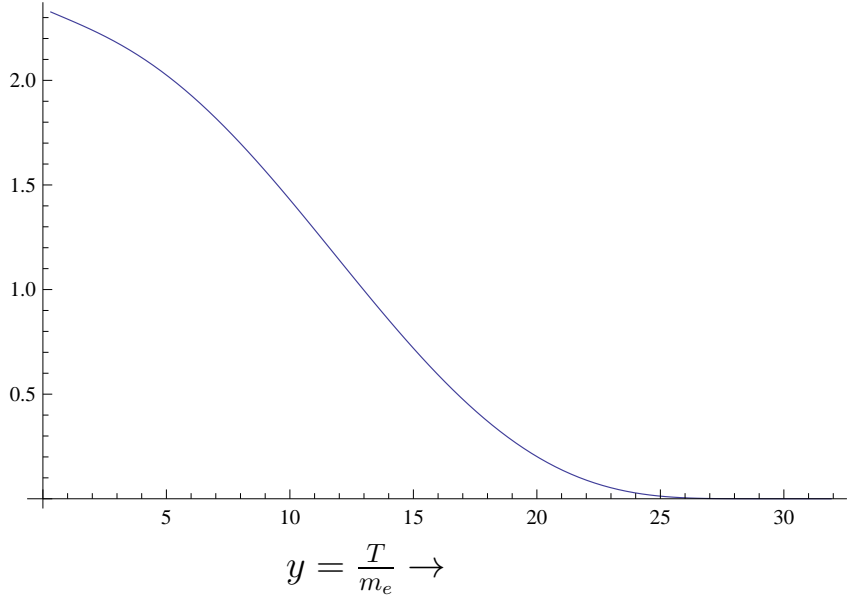}\\
{\hspace{-1.5cm}$ y=\frac{T}{m_e}\rightarrow$ }
\caption{The function $\prec\left(\frac{d\tilde{\sigma}}{dy}\right)\succ$ as a function of the electron energy in units of $m_e$ in the case of Boron neutrinos.
 \label{fig:dsigmadyBe}}
\end{center}
\end{figure}

\begin{figure}
\begin{center}
\includegraphics[width=0.8\textwidth]{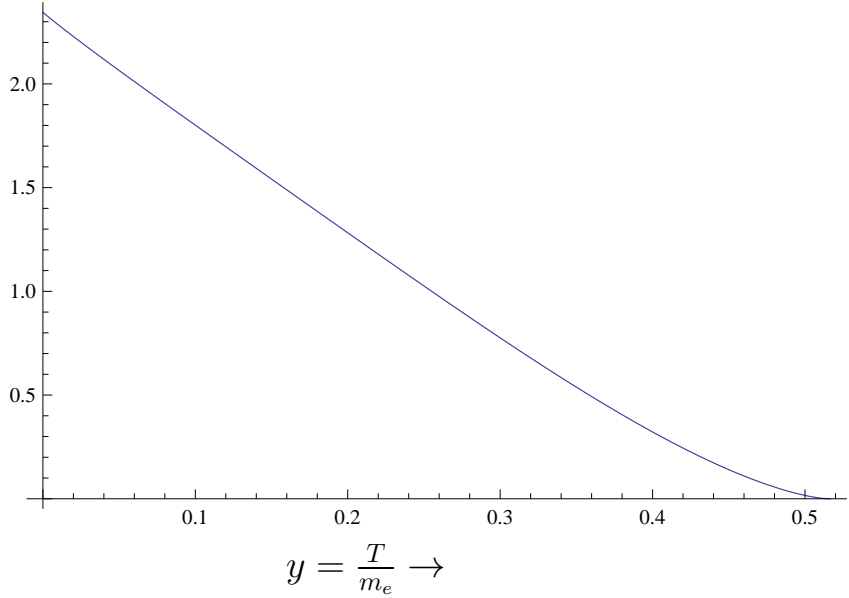}\\
{\hspace{-1.5cm}$ y=\frac{T}{m_e}\rightarrow$ }
\caption{The function $\prec\left(\frac{d\tilde{\sigma}}{dy}\right)\succ$ as a function of the electron energy in units of $m_e$ in the case of pp neutrinos neglecting the electron binding energy.
 \label{fig:dsigmadyppe0}}
\end{center}
\end{figure}

\begin{figure}
	\begin{center}
\includegraphics[width=0.8\textwidth]{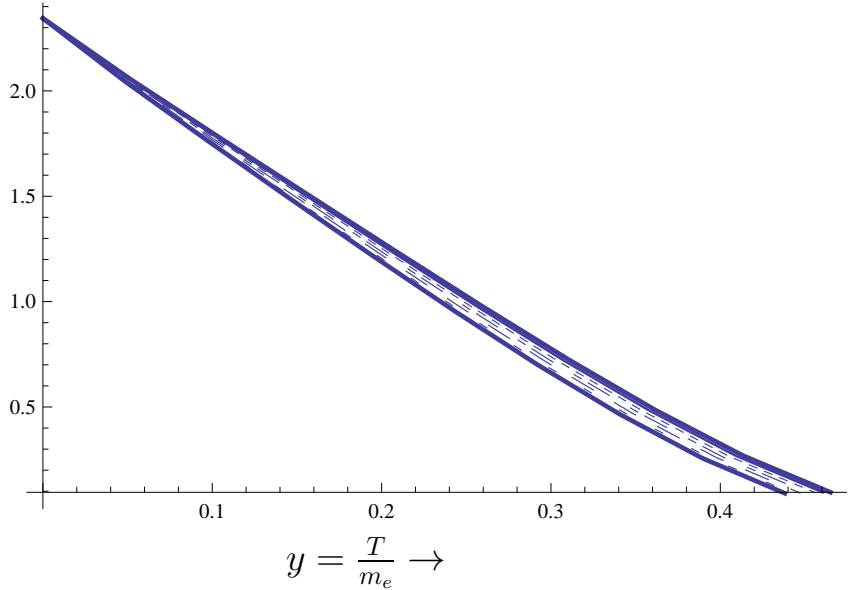}\\
{\hspace{-1.5cm}$ y=\frac{T}{m_e}\rightarrow$ }
\caption{The function $\prec\left(\frac{d\tilde{\sigma}}{dy}\right)\succ$ as a function of the electron energy in units of $m_e$ in the case of pp neutrinos for various binding energies, as e.g. are found in the Xe nucleus.
 \label{fig:dsigmadyppeb}}
\end{center}
\end{figure}

The dimensionless expression  $\prec\left(\frac{d\tilde{\sigma}}{dy}\right)\succ$, related to the average differential cross section (see Eqs \ref{Eq:avedsdy1} and \ref{Eq:avedsdy2}),  is exhibited in \ref{fig:dsigmadyBe} in the case of Boron neutrinos. One sees that the average cross section has a maximum at zero electron energy (all neutrinos  contribute) and decreases with electron energy, because only a fraction of neutrinos contribute, since the required minimum neutrino energy is increasing with electron energy (see Eq. (\ref{Eq:xmin})). 
Integrating over the  spectrum we find that the the expected cross section, $\sigma_{B}$, in units of $ws$ (see Eq: \ref{Eq:weakscale}),  is $\approx 23$.
Similarly  the case of electrons produced by   pp neutrinos is presented in Fig. \ref{fig:dsigmadyppe0}.
As we have mentioned, however, the binding energy of the electrons in the case of low energy the pp neutrinos cannot be a priori neglected.
Thus we find that $x_{min}$ in Eq. \ref{Eq:sigmatot} becomes:
$$
x_{min}=\frac{1}{2 \left(b^2+2 b y-2 y\right)}$$ $$
\left (b^3+3 b^2 y+b^2 \sqrt{y (y+2)}+2 b y^2-2 b y+2 b y \sqrt{y (y+2)}\right .$$ $$ \left . -2 y^2-2
   y \sqrt{y (y+2)} \right ) $$
	where $b$ is the electron binding energy. The maximum electron energy is also affected a little bit.
	The corresponding electron spectrum is now  shown in Fig. \ref{fig:dsigmadyppeb}. We see that the effect is small. In practice it is even smaller since only the K-electrons in a potential target like Xe have a binding energy, which is as much as  6$\%$ of the electron mass.
Integrating over the the spectrum we find that the the expected cross section, $\sigma_{pp}$, in units of $ws$ (see Eq: \ref{Eq:weakscale}), is between 0.53 and 0.49 depending on the binding energy.
\section{Some results}
The number of the expected events  is
\beq
N_{ev,i}=\Phi_i N_e \sigma_i  sc, N_e=N_A Z,\, i=\mbox{pp and B respectively}
\eeq
where $Z$ is the atomic number and $N_A$ is the number of Nuclei in the target. Using a Kg of Xe target we find that 
 $$N_A = 1/( 127\times 1.6726219 \times10 ^{-27})=4.7\times10^{24}$$
Using this input we get:
\beq
N_{ev,pp}=0.53,\,N_{ev,B}=2 \times 10^{-4},\mbox{ events per Kg-y}
\eeq
The higher cross section for boron neutrinos, is no match for the much higher flux of pp neutrinos.
Thus the Boron neutrinos are harmless, but the pp neutrinos may be a serious background. 

Estimates of the rate for WIMP-electron scattering for light WIMPs are of order 0.5   events per Kg-y for atomic electrons\cite{VMEK16}, close to the above background. The expected rate is  suppressed significantly if one includes binding energy effects. Such effects have not always been properly considered, but, in fact, the situation is similar with that  appearing in the standard inelastic WIMP-nucleus scattering \cite{VAKPSST16}. In the latter case only states with excitation energy $E_x$ satisfying the condition 
\beq
E_x\leq \frac{1}{2} \mu_r \upsilon_{esc}^2,
\eeq
can be populated. $\mu_r$ is the reduced mass of the WIMP-nucleus system and $\upsilon_{esc}$ is the WIMP escape velocity which is $\approx 2 \times 10^{-6}c$  This yields the  approximate rule $$ \mbox{ 20 GeV WIMP }\Leftrightarrow E_x\leq \mbox{ 40 keV}  $$
A similar situation prevails in the case of WIMP-electron scattering. Now energy is lost to eject a bound electron:
\beq
b\leq \frac{1}{2} \mu_r \upsilon_{esc}^2=\frac{1}{2} m_e \frac{1}{1+m_e/m_{\chi}}\upsilon_{esc}^2,
\eeq
$$ \mbox{ 0.5 MeV WIMP }\Leftrightarrow  b\leq \mbox{ 0.5 eV} $$
This means that few electrons are available to the scattering, i.e. $Z_{eff}<Z$. Furthermore, even if an electron is ejected, the kinematics is unfavorable. Thus in the case of $b=0.5$ eV the minimum electron energy is 0.05 eV, while the maximum is  between 0.5 and 3 eV for a WIMP with mass between $m_e$ and $50m_e$.

\begin{figure}
\begin{center}
\includegraphics[width=0.8\textwidth]{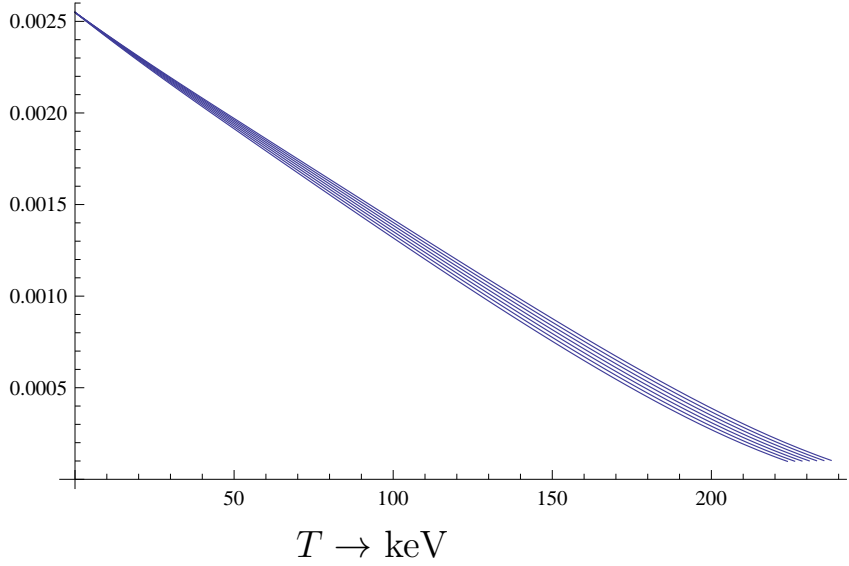}\\
{\hspace{-1.5cm}$ T\rightarrow$ keV}
\caption{The event rate per kg-y/keV for electrons produced by pp solar neutrinos as a function of the electron energy in keV in the case of a Xe target. The band corresponds to the range of the electron  binding energy  in the case of this target.
 \label{fig:dRdEkeVppe}}
\end{center}
\end{figure}

 At this point the spectrum of electrons arising from pp neutrinos, similar to that of Fig. \ref{fig:dsigmadyppeb}, but  in the experimentally useful  units, (Kg-y/keV), is presented in Fig. \ref{fig:dRdEkeVppe}. The area under the curve yields now  the total number of background events.

It is well known that the rates involving electrons, background as well as those involved in WIMP searches, are expected to be largely enhanced if the Fermi function is included, since  the produced electrons have low energy. The relative importance of the background, however,  is not going to be affected by this function and  we will not elaborate further here.\\

\section{Conclusions}
In the present paper we studied the solar neutrino  background which may affect the  search for light dark matter candidates in the case they employ electron detectors. We find that the background due to the low energy pp neutrinos is serious at the level of 0.5 events per Kgr-y.

{\bf Acknowledgments:}\\
This work was partially supported by the University of Adelaide and by the Australian Research Council through the ARC Centre of Excellence for Particle
Physics at the Terascale, CE110001104, and  DP150103101.

\end{document}